\documentclass[twocolumn,showpacs,preprintnumbers,amsmath,amssymb]{revtex4}

\usepackage{graphicx}
\usepackage{dcolumn}
\usepackage{bm}

\begin{document}
\def\vv{{\hbox{\boldmath $v$}}}
\def\bsb{{\hbox{\boldmath $\beta$}}}

\title{Two-particle interferometry for the sources undergoing first-order
QCD phase transition in high energy heavy ion collisions}

\author{Hong-Jie Yin}
\author{Jing Yang}
\author{Wei-Ning Zhang\footnote{wnzhang@dlut.edu.cn}}
\affiliation{School of Physics and Optoelectronic Technology,
Dalian University of Technology, Dalian, Liaoning 116024, China }
\author{Li-Li Yu}
\affiliation{College of Science, Nanjing Forestry University, Nanjing,
Jiangsu 210037, China}
\date{\today}

\begin{abstract}
We investigate the two-particle interferometry for the particle-emitting
sources which undergo the first-order phase transition from the quark-gluon
plasma with a finite baryon chemical potential to hadron resonance gas.
The effects of source expansion, lifetime, and particle absorption on the
transverse interferometry radii $R_{\rm out}$ and $R_{\rm side}$ are
examined.  We find that the emission durations of the particles become
large when the system is initially located at the boundary between the
mixed phase and the quark-gluon plasma.  In this case, the difference
between the radii $R_{\rm out}$ and $R_{\rm side}$ increases with the
transverse momentum of the particle pair significantly.  The ratio of
$\sqrt{R_{\rm out}^2 -R_{\rm side}^2}$ to the transverse velocity of
the pair is an observable for the enhancement of the emission duration.
\end{abstract}
\maketitle

\section{Introduction}
One of the important issues of high energy heavy ion collisions is to
find and quantify the quantum chromodynamics (QCD) phase transition
between the quark-gluon plasma (QGP) at higher energy density and the
hadron gas at lower energy density.  The initial systems produced
in the heavy ion collisions at the higher energies of the Super Proton
Synchrotron (SPS) and Relativistic Heavy Ion Collider (RHIC) and the
energy of the Large Hadron Collider (LHC) have high temperature and
near-zero baryon chemical potential.  Lattice QCD calculations have
shown that the transition at the vanishing baryon chemical potential is
a crossover \cite{Aok06}.  However, it is predicted that this crossover
will become a first-order phase transition at intermediate temperatures
and high baryon chemical potentials \cite{{Ant03,Hat03,Fod04,Gav05,
Asa08,Eji08}}.  Recently, the search for the evidences of the first-order
phase transition and location of its critical end point have attracted
special attention, for instance, the RHIC and SPS low energy programs
\cite{Ste0608,STAR10,NA49-06,NA61-09,NA49-NA61-11} and the project
of the future Facility for Antiproton and Ion Research (FAIR) at
GSI \cite{Fri06,Pet06Ros07,Hen08,HohJoh09}.

Two-particle Hanbury-Brown-Twiss (HBT) interferometry is a useful
tool for detecting the space-time structure of particle-emitting
sources in high energy heavy ion collisions
\cite{Won94,Wie99,Wei00,Lis05}.  For the first-order phase
transition there is a mixed phase of the QGP and hadron gas.
In the absence of pressure gradient, a slow-burning fireball is
expected when the initial system is at rest in the mixed phase,
and this may lead to a considerable time-delay of the system
evolution \cite{Pra861,Ber89,Hun95,Ris96,Sof02,Zsc02}.  It is
therefore of interest to probe the time-delay for the first-order
phase transition by HBT interferometry.

In Ref. \cite{Yul08} an HBT analysis technique with quantum transport
of the interfering pair (QTIP) is developed.  It takes into account
the effects of resonance decay and multiple scattering of pions in
the sources.  In Ref. \cite{Yul10}, this HBT technique is used to
investigate the source radius and lifetime for the spherical systems
evolving hydrodynamically with the first-order phase transition.
In this study we use the relativistic hydrodynamics in (2+1) dimension
to describe the particle-emitting sources which undergo the first-order
phase transition.  We investigate the HBT radii $R_{\rm out}$,
$R_{\rm side}$, and $R_{\rm long}$ \cite{Ber88,Pra90} for the
hydrodynamic sources using the HBT interferometry with the QTIP
technique.  The results indicate that the ratio of $\Delta R_{\rm os}
=\sqrt{R_{\rm out}^2 -R_{\rm side}^2}$ to the transverse velocity of
the particle pair $v_{_{\rm KT}}$ is sensitive to the emission duration
of the source.  It is large when the system is initially located at the
boundary of the mixed phase and the QGP (soft point).  As compared to
pion HBT interferometry kaon HBT interferometry may present more clearly
the source space-time geometry at the emission, because kaons (for
instance ${\rm K}^+$) can escape easily from the system after their
production.  By comparing the results of the two-pion and two-kaon HBT
analyses, we find that the particle absorptions and the large
expansion velocities of the sources after hadronization may change the
pion HBT radii as functions of the transverse momentum of the particle
pair.  However, the large values of the ratio $\Delta R_{\rm os}/
v_{_{\rm KT}}$ for the soft point of the first-order phase transition
can be observed in both of two-pion and two-kaon HBT measurements.

The paper is organized as follows.  In section II we present briefly the
description for the relativistic hydrodynamics in cylindrical coordinate
frame.  We describe the model of the equation of state (EOS) of first-order
phase transition used in our calculations.  The adiabatic cooling paths and
the space-time evolution of the systems are also discussed in this section.
In section III we perform the two-pion and two-kaon HBT analyses, with the
QTIP technique, for the hydrodynamic particle-emitting sources for the
initial conditions of the QGP and the soft point of the first-order phase
transition.  The effects of source expansion, lifetime, and particle
absorptions on transverse HBT radii are investigated.  On the basis of the
investigations, we introduce an observable to probe the long lifetime of
the source for the initial conditions of the soft point.  Finally, the
summary and conclusions are presented in section IV.

\section{Hydrodynamical evolution with first-order phase transition}

\subsection{Relativistic hydrodynamic equations in cylindrical frame}
The dynamics of ideal fluid in high energy heavy ion collisions is
defined by the local conservations of energy-momentum and net
charges \cite{Ris98,Kol03}.  The continuity equations of the
conservations of energy-momentum, net baryon number, and entropy
are
\begin{eqnarray}
\label{Tuv}
\partial_{\mu} T^{\mu\nu}(x)=0 \,,
\end{eqnarray}
\begin{eqnarray}
\label{bnc}
\partial_{\mu}j^{\mu}_b (x)=0 \, ,
\end{eqnarray}
\begin{eqnarray}
\label{sc}
\partial_{\mu}j^{\mu}_s (x)=0 \, ,
\end{eqnarray}
where $x$ is the space-time coordinate of a thermalized fluid
element in the source center-of-mass frame, $T^{\mu\nu}(x)$ is the
energy momentum tensor of the element, $j^\mu_b(x)=n_b(x)u^\mu$ and
$j^\mu_s(x)= s(x)u^\mu$ are the four-current-density of baryon and
entropy ($n_b$ and $s$ are the baryon density and entropy density),
and $u^\mu=\gamma(1,\vv)$ is the four-velocity of the fluid element.
The energy momentum tensor $T^{\mu\nu}(x)$ is given
by \cite{Ris98,Kol03}
\begin{equation}
\label{tensor}
T^{\mu \nu} (x) = \big [ \varepsilon(x) + p (x) \big ] u^{\mu}(x)
u^{\nu}(x) - p (x) g^{\mu\nu} \, ,
\end{equation}
where $p$ and $\varepsilon$ are the pressure and energy density of
the fluid element, and $g^{\mu\nu}$ is the metric tensor.

In the cylindrical coordinate $(t, \rho, \phi, z)$ frame,
$g^{\mu\nu}={\rm diag} (1, -1, -\rho^{-2}, -1)$.  The conservation
Eqs. (\ref{Tuv}) -- (\ref{sc}) can be expressed as
\begin{equation}
\label{EE}
\partial_t E +\partial_{\rho} [(E+p)v^{\rho}] +\partial_z [(E+p)v^z]
= -\frac{v^{\rho}}{\rho} (E+p) ,
\end{equation}
\begin{equation}
\label{MR}
\partial_t M^{\rho} +\partial_{\rho} (M^{\rho} v^{\rho} + p)+
\partial_z( M^{\rho} v^z) = -\frac{v^{\rho}}{\rho} M^{\rho} ,
\end{equation}
\begin{equation}
\label{MZ}
\partial_t M^z + \partial_{\rho} (M^z v^{\rho}) + \partial_z
(M^z v^z + p\,) = -\frac{v^{\rho}}{\rho} M^z ,
\end{equation}
\begin{equation}
\label{NB}
\partial_t N_b +\partial_{\rho} (N_b v^{\rho}) +\partial_z
(N_b v^z) = -\frac{v^{\rho}}{\rho} N_b ,
\end{equation}
\begin{equation}
\label{NS}
\partial_t N_s +\partial_{\rho} (N_s v^{\rho}) +\partial_z
(N_s v^z) = -\frac{v^{\rho}}{\rho} N_s ,
\end{equation}
where $E \equiv T^{00}$, $M^{\rho} \equiv T^{0{\rho}} =
T^{{\rho} 0}$, $M^z \equiv T^{0z} = T^{z0}$, $N_b \equiv j^0_b
= n_b \gamma$, and $N_s \equiv j^0_s = s \gamma$.

\subsection{Equation of state}
In the equations of motion (\ref{EE}) -- (\ref{NS}), there are
$\varepsilon$, $p$, $v^{\rho}$, $v^z$, $n_b$, and $s$ six unknown
functions.  In order to obtain the solution of the equations of
motion, we need an equation of state (EOS), $p(\varepsilon, n_b,
s)$, which gives a relation for $p$, $\varepsilon$, $n_b$, and $s$.
In our model the QGP phase is described by a perfect gas of gluons,
$u$, $d$, $s$ quarks, and antiquarks, with the constant vacuum
energy $B$ associated with QCD confinement \cite{Ton03}.  The
pressure, energy density, and the conserved charge density in
the QGP phase are given by
\begin{equation}
\label{eq-q-eosp}
p^Q = \sum_i p_i(T,\mu_i) - B \,,
\end{equation}
\begin{equation}
\label{eq-q-eose}
\varepsilon^Q = \sum_i \varepsilon_i(T,\mu_i) + B
\,,
\end{equation}
\begin{equation}
\label{eq-q-eosn}
n_A^Q =  \sum_i A_i \  n_i(T,\mu_i) \,,
\end{equation}
where $p_i(T,\mu_i)$, $\varepsilon_i(T,\mu_i)$, and $n_i(T,\mu_i)$
are the pressure, energy density, and number density of particle
species $i$ in the perfect gas with temperature $T$ and chemical
potential $\{\mu_i\}$, $A_i$ is the conserved charge number of the
particle species $i$.  In our calculations we use the quark masses
$m_u=m_d=5$ MeV, $m_s=150$ MeV and the bag constant
$B=(235~\mbox{MeV})^4$ \cite{Ton03}.

For the hadronic phase we adopt the excluded volume model
\cite{Ris91,Hun98,Ton03} and consider the particles $\pi$, $K$, $N$,
$\Lambda$, $\Sigma$, $\Delta$, and their antiparticles in the model.
The pressure, energy density, and the conserved charge density in
the hadronic phase are given by \cite{Ris91,Hun98,Ton03}
\begin{eqnarray}
\label{eq-h-eosp} p^H = \sum_i p_i(T,\tilde\mu_i)\,,
\end{eqnarray}
\begin{eqnarray}
\label{eq-h-eose} \varepsilon^H = {
\sum_i\varepsilon_i(T,\tilde\mu_i) \over 1 + V_0\sum_i
{n_i(T,\tilde\mu_i) } } \,,
\end{eqnarray}
\begin{eqnarray}
\label{eq-h-eosn} n_A^H = {\sum_i A_i\, n_i(T,\tilde\mu_i) \over 1 +
V_0\sum_i {n_i(T,\tilde\mu_i) } } \,,
\end{eqnarray}
where
\begin{eqnarray}
\tilde\mu_i = \mu_i - V_0\, p^H\,,
\end{eqnarray}
$V_0=(1/2)(4\pi/3)(2a)^3$ is the excluded volume which is assumed to
be the same for all hadrons with $a=0.5$ fm \cite{Ton03}.

For the first-order phase transition, there are Gibbs relationships
in the mixed phase of the QGP and hadron gas.  We have $T^Q=T^H$,
$\mu_{N,\Delta}=3\mu_u$, $\mu_{\Lambda,\Sigma} =2\mu_u+\mu_s$,
$\mu_{\pi^+,\pi^0,\pi^-}=0$, $\mu_{K^+,K^0}=\mu_u-\mu_s$, ..., and
\begin{eqnarray}
\label{eq-m-eosp} p^M=p^Q(T,\mu_u,\mu_s)=p^H(T,\mu_u,\mu_s) \,,
\end{eqnarray}
\begin{eqnarray}
\label{eq-m-eose}
\varepsilon^M = \alpha \,
\varepsilon^Q(T,\mu_u,\mu_s ) + (1-\alpha ) \,
\varepsilon^H(T,\mu_u,\mu_s ) \,,
\end{eqnarray}
\begin{eqnarray}
\label{eq-m-eosn}
n_A^M = \alpha \, n_A^Q(T,\mu_u,\mu_s ) +
(1-\alpha ) \, n_A^H(T,\mu_u,\mu_s ) \,,
\end{eqnarray}
where $\mu_u$ and $\mu_s$ are the chemical potentials of $u$ and $s$
quarks, and $\alpha = V_Q / V$ is the fraction of the volume
occupied by the plasma phase.  The boundaries of the coexistence
region are found by putting $\alpha = 0$ (the hadron phase boundary)
and $\alpha = 1$ (the plasma boundary).

Using the thermodynamical relations of mixed gas one can get the
entropy densities $s$ and other thermodynamical quantities, in the
QGP, hadronic, and mixed phases from Eqs. [(\ref{eq-q-eosp}) --
(\ref{eq-q-eosn})], [(\ref{eq-h-eosp}) -- (\ref{eq-h-eosn})], and
[(\ref{eq-m-eosp})--(\ref{eq-m-eosn})], and get numerically the EOS
with the first-order phase transition.

\subsection{Adiabatic paths}
\begin{figure}
\includegraphics[angle=0,scale=0.50]{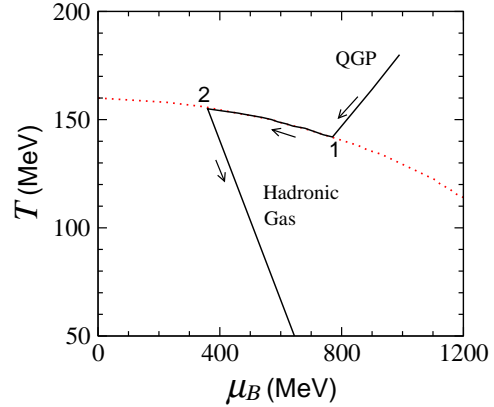}
\caption{\label{fig:adiabat} (Color online) The adiabatic paths for the
system evolving with the EOS of the first-order phase transition.  The
dotted line is the transition curve between the QGP and hadron gas.
The mixed phase is on the transition curve from the end point of the QGP
branch (point 1) to the beginning of the hadronic branch (point 2). }
\end{figure}

For perfect fluid, the entropy and baryon number of the system are
conserved during evolution.  So the ratio of the densities $n_b$ and
$s$, $n_b/s$, is a constant.  In the calculations we take $n_b/s=0.06$
which corresponds to the incident energy about 30 $A$GeV \cite{Iva06}.
The solid lines in Fig. \ref{fig:adiabat} show the adiabatic cooling
paths for the system evolving with the EOS of the first-order phase
transition.  The dotted line is the transition curve between the QGP
and hadron gas.  The mixed phase is on the transition curve from the
end point of the QGP branch (point 1) to the beginning of the hadronic
branch (point 2).  The non-trivial zigzag shape of the trajectory indicates
that the system has a re-heating in the mixed phase \cite{Sub86,Hun98}.
The reason is that at a certain point ($T, \mu$) on the phase-transition
curve, the number of degrees of freedom, and hence the specific entropy,
is larger in the QGP phase than which in the hadronic phase.  The
temperature must increase during hadronization to conserve both the
total entropy and baryon number simultaneously \cite{Sub86}.

\begin{figure}
\includegraphics[angle=0,scale=0.50]{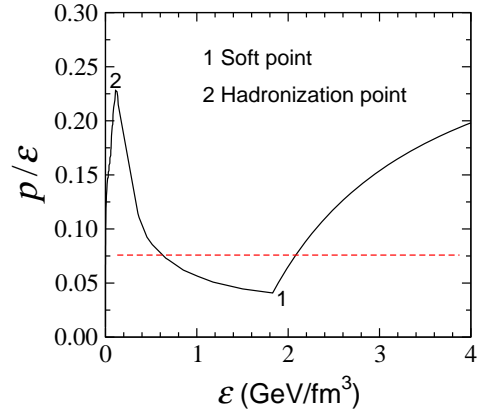}
\caption{\label{fig:pe} (Color online) The ratio of the system pressure to
energy density, $p/\varepsilon$.  It has the minimum at the soft point 1 and
reaches its maximum at the hadronization point 2.  The ratio retains the
values smaller than 0.075 in the $\varepsilon$ regain 0.6 -- 2.1 GeV/fm$^3$}
\end{figure}

In Fig. \ref{fig:pe} we show the thermodynamical quantity, $p/\varepsilon$,
as a function of $\varepsilon$ for the system.  The ratio $p/\varepsilon$
reaches the minimum at the boundary between the QGP and mixed phase,
$\varepsilon=\varepsilon^{\rm MQ}=1.83$ GeV/fm$^3$.  It is so called the
soft point of the first-order phase transition.  At the boundary between
the mixed phase and hadronic gas, the ratio reaches its maximum.  It is named
hadronization point.  One can see that the ratio retains the values smaller
than 0.075 in the $\varepsilon$ regain 0.6 -- 2.1 GeV/fm$^3$.

\subsection{System evolution}

\begin{figure}
\vspace*{-5mm}
\includegraphics[angle=0,scale=0.42]{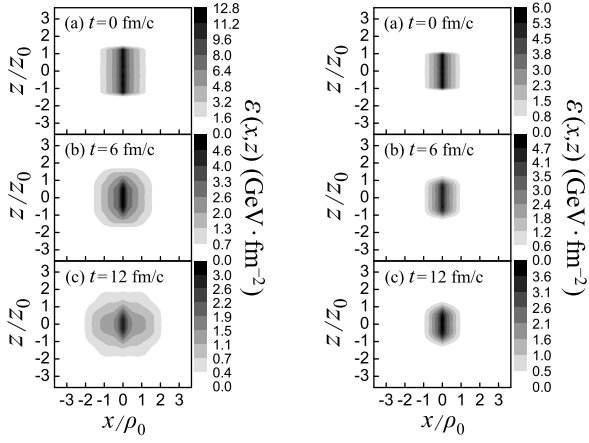}
\vspace*{-23mm}
\caption{\label{fig:evolT} The two-dimension energy density,
$\varepsilon(x,z)$, for the systems at the time $t=$0, 6, and 12 fm/c.
The left panels are for the system initially in the QGP.  The
right panels are for the system initially at the soft point.}
\vspace*{5mm}
\end{figure}

\begin{figure}
\includegraphics[angle=0,scale=0.62]{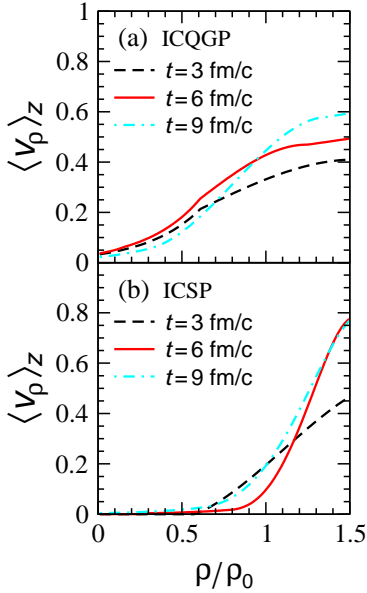}
\caption{\label{fig:veloT} (Color online) The average transverse velocity
for the systems initially in the QGP (a) and at the soft point (b). }
\end{figure}

Using the Sod's operator splitting and RHLLE method
\cite{Sod77,Sch93,Ris95,Ris98}, we can obtain the system evolution
by solving the hydrodynamical equations (\ref{EE})--(\ref{NS})
with the EOS of the first-order phase transition.  Because the
heavy ion collisions are full stopped at the energy considered,
we assume the system is initially at rest within a cylinder in the
beam direction ($z$-direction) with the transverse and longitudinal
radii $\rho_0$ and $z_0$.  In Fig. \ref{fig:evolT}, we show the
two-dimension energy density, $\varepsilon(x,z)= \int \varepsilon
(x,y,z)dy$, for the systems at the time $t=$0, 6, and 12 fm/c.  The
left and right panels are for the systems which are initially located
in the QGP phase ($T_0^{\rm QGP}=180$ MeV, $\varepsilon_0^{\rm QGP}=
4.12$ GeV/fm$^3$, $\mu_{B0}^{\rm QGP} = 990$ MeV) and at the soft
point ($T_0^{\rm SP}=142$ MeV, $\varepsilon_0^{\rm SP}=
\varepsilon^{\rm MQ}=1.83$ GeV/fm$^3$, $\mu_{B0}^{\rm SP}=780$ MeV).
It can be seen that the energy density for the system with the initial
conditions of the soft point (ICSP) decreases more slowly with time
than that for the system with the initial conditions of the QGP
(ICQGP).  Because there are not the initial velocity and pressure
gradient in the mixed phase, the expansion of the system is slow.
Figure \ref{fig:veloT} (a) and (b) show the average transverse
velocity, $\langle v_{\rho}(\rho)\rangle_z =\int v_\rho (\rho,z)dz$,
for the systems with ICQGP and ICSP, respectively.
For ICQGP, the velocity increases rapidly from zero at the beginning
($t\leq 3$ fm/c), and still increases with time during $3<t<6$ fm/c.
Because there is larger gradient of pressure on the edge of the system,
the velocity increase more rapidly around $\rho \sim \rho_0$.  At $t=9$
fm/c, the decrease of the velocity near the center of the system is due
to the blast-wave expansion which leads to a void in the center region.
For ICSP, the velocity retains zero in the center region of the system
even at a larger time because there is not pressure gradient in this case.
In our calculations, the initial sizes for the system with ICQGP are
taken to be $\rho_0=z_0=$4.0 fm.  The initial sizes for the system with
ICSP are taken to be $\rho_0=z_0=4.0\times(\varepsilon_0^{\rm QGP}/
\varepsilon^{\rm MQ})^{1/3}=$5.2 fm.

\section{HBT interferometry with quantum transport of the interfering pair}

\subsection{Formulas of correlation function}

The two-particle HBT correlation function $C(k_1,k_2)$ is defined as the
ratio of the two-particle momentum distribution $P(k_1,k_2)$ to the
product of the single-particle momentum distribution $P(k_1) P(k_2)$,
\begin{equation}
\label{Ck1k2}
C(k_1,k_2)=\frac{P(k_1,k_2)}{P(k_1)P(k_2)}.
\end{equation}
Using the quantum probability amplitudes in a path-integral formalism
\cite{Yul08}, $P(k_i)$ ($k_i=(E_i, {\bf k}_i),~ i=1,2$) and $P(k_1,k_2)$
can be expressed as
\cite{Won03,Won04,Won05,Zha04c,Yul08}
\begin{eqnarray}
\label{pk1} P(k_i)=\int\!d^4x \,\rho(x)e^{-2\,{\cal I}{m}\,{\bar
\phi}_s(x\kappa \to k_i)} |A(x\kappa)|^2\,,
\end{eqnarray}
\begin{eqnarray}
\label{pk12}
P(k_1,k_2)\!\!&=&\!\!\!\!\int\!\!d^4x_1 d^4x_2\,e^{-2{\cal I}{m}
{\bar\phi}_s(x_1\kappa_1\to k_1)} e^{-2{\cal I}{m} {\bar\phi}_s
(x_2\kappa_2 \to k_2)}\nonumber\\
&&\!\!\times\rho(x_1) \rho(x_2)|\Phi(x_1 x_2; k_1 k_2)|^2 ,
\end{eqnarray}
where $\rho(x)$ is the four-dimension density of the
particle-emitting source, $A(x\kappa)$ is the amplitude for producing
a particle at $x$ with momentum $\kappa$, $e^{-2{\cal I}{m}{\bar\phi}_s
(x\kappa \to k)}$ is the absorption factor due to the multiple scattering
when the particle propagating in the source, and $\Phi(x_1 x_2;k_1 k_2)$
is the wave function for the two identical bosons,
\begin{eqnarray}
\Phi(x_1 x_2; k_1 k_2)=&&\nonumber\\
&&\hspace*{-25mm}\frac{1}{\sqrt{2}} \{ \bar A(x_1 \kappa_1,k_1)
\bar A(x_2 \kappa2,k_2) e^{ik_1\cdot x_1+ik_2\cdot k_2}\nonumber\\
&&\hspace*{-23mm}+ \{ \bar A(x_1 \kappa_2',k_2) \bar A(x_2 \kappa_1',
k_1) e^{ik_1\cdot x_1+ik_2\cdot k_2} \},
\end{eqnarray}
\begin{eqnarray}
\hspace*{-10mm}\bar A(x\kappa,k)=A(x\kappa)e^{i\delta_{\rm mf}(x\kappa
\to k)},
\end{eqnarray}
where $\delta_{\rm mf}(x\kappa \to k)$ is a phase arising from the
source collective expansion, which can be described by a long-range
density-dependent mean-field \cite{Won05,Yul08}.

In our HBT calculations, the identical kaons (for instance K$^+$) are
assumed to freeze out directly at the hadronization.  So, the absorption
factor $e^{-2{\cal I}{m}{\bar\phi}_s}$ is 1 and $\delta_{\rm mf}=0$.
The final identical pions (for instance $\pi^+$) include the primary
pions emitted at the hadronization and the secondary pions from the
``excited-state" particle decays during the system evolving in hadronic
phase until to the thermal freeze-out.  The four-dimension density of
the pion source can be expressed as \cite{Beb92,Hir02,Yul08}
\begin{equation}
\rho(x)=n_{\pi}(x)\delta(t-\tau^h)+\sum_{j\ne \pi}D_{j\rightarrow
\pi} n_j(x)\,,
\end{equation}
where $n_i(x)$ and $\tau^h$ are the particle number density and the
hadronization time in local frame, $D_{j\rightarrow \pi}$ is the
product of the decay rate in time and the fraction of the decay. For
example, $D_{\Delta \rightarrow \pi}= \Gamma_{\Delta}\times \frac{1}
{3}$ and $D_{\pi^0\pi^0 \rightarrow \pi^+\pi^-}=v_r n_{\pi} \sigma
(\pi^0\pi^0 \rightarrow \pi^+\pi^-) \times 1$, where $v_r$ is the
relative velocity of the two colliding pions and the cross section
$\sigma(\pi^0\pi^0 \rightarrow \pi^+\pi^-)$ is equal to the absorption
cross section of $\pi^+\pi^- \rightarrow \pi^0\pi^0$ \cite{Yul08}.

When a pion propagating in the source it will subject to multiple
scattering with the medium particles in the source.  The absorption
factor due to the multiple scattering in Eqs. (\ref{pk1}) and
(\ref{pk12}) can be written as \cite{Won03,Won04,Won05,Zha04c,Yul08}
\begin{eqnarray}
\label{absf} e^{-2\,{\cal I}{m}\,{\bar \phi}_s(x)}\!=\exp\Bigg[\!-\!
\!\int_{x}^{x_f}\!\!\Big({\sum_i}'\sigma_{\rm abs}(\pi i)\ n_i(x')\Big)
d\ell(x')\Bigg],\nonumber\\
\end{eqnarray}
where $\sum'_i$ means the summation for all medium particles except
for the test pion along the propagating path $d\ell(x')$,
$\sigma_{\rm abs}(\pi i)$ is the absorption cross section of the
pion with the particle species $i$ in the medium, and $x_f$ is
the freeze-out coordinate.  In calculations we only consider the
dominant absorption processes for the identical pions, for example
the reactions of $\pi^+\pi^- \to \pi^0 \pi^0$ and $\pi^+ N
\rightarrow \Delta$ for $\pi^+$, as in Ref. \cite{Yul08}.  The pion
freeze-out temperature is taken to be 110 MeV, which corresponds to
the energy density $\varepsilon_f=$45 MeV/c \cite{Cle99}.

In the HBT analysis, we use the Bertsch-Pratt components of the
relative momentum $q=|{\bf k}_1 -{\bf k}_2|$ of the identical particle
pair \cite{Ber88,Pra90}, $q_{\rm side}$, $q_{\rm out}$, and $q_{\rm long}$
as variables.  The correlation function $C_K(q_{\rm side},q_{\rm out},
q_{\rm long})$ are constructed from $P(k_1,k_2)$ and $P(k_1)P(k_2)$ by
summing over ${\bf k}_1$ and ${\bf k}_2$ for the $(q_{\rm side},q_{\rm
out},q_{\rm long})$ bins in a certain $K_T=\frac{1}{2}|{\bf k}_1 -{\bf
k}_2|_T$ region.  The HBT radii $R_{\rm side}(K_T)$, $R_{\rm out}(K_T)$,
and $R_{\rm long}(K_T)$ are obtained by fitting the correlation
functions with the parametrized formula
\begin{eqnarray}
C_K(q_{\rm side},q_{\rm out},q_{\rm long})&=&1+{\lambda} \exp[-q_{\rm
side}^2R_{\rm side}^2(K_T) \hspace*{5mm}\nonumber\\
&&\hspace*{-15mm}-q_{\rm out}^2R_{\rm out}^2(K_T) -q_{\rm long}^2
R_{\rm long}^2(K_T)]\,,
\end{eqnarray}
in the longitudinal comoving system (LCMS).  Here $\lambda$ is called
the chaotic parameter.

\subsection{Results for hydrodynamic sources}

\begin{figure}
\includegraphics[angle=0,scale=0.68]{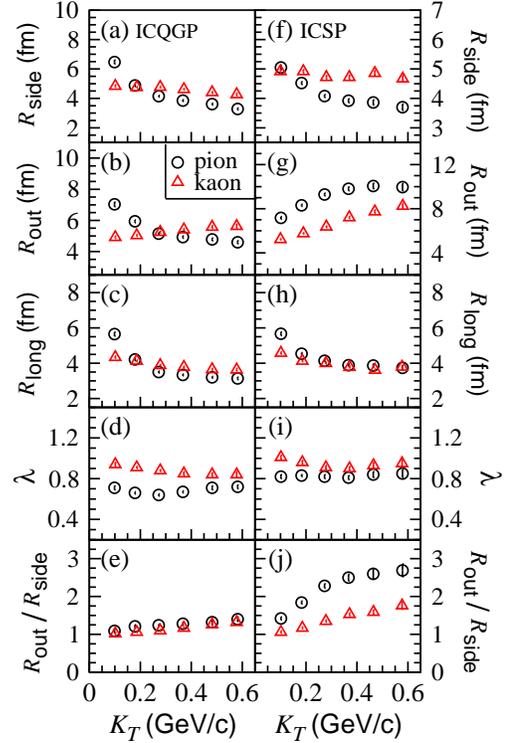}
\caption{\label{fig:HBTr06} (Color online) The two-pion and two-kaon
HBT results for the hydrodynamic sources for ICQGP and ICSP. }
\end{figure}

In Fig. \ref{fig:HBTr06} we show the two-pion and two-kaon HBT results
for the hydrodynamic sources for ICQGP and ICSP.  It can be seen that
there is much difference for the two-pion HBT radius $R_{\rm out}$ as
functions of $K_T$ for the two kinds of sources.  One decreases with
$K_T$, and another almost increase with $K_T$.  When the system is
initially located at the soft point (ICSP case), the results of
$R_{\rm out}$ are much larger than those of $R_{\rm side}$ at larger
$K_T$, and the ratio $R_{\rm out}/R_{\rm side}$ increases with $K_T$
significantly.  As compared to the pion HBT radii the kaon HBT
radii exhibit more moderate changes with $K_T$.

\begin{figure}
\includegraphics[angle=0,scale=0.68]{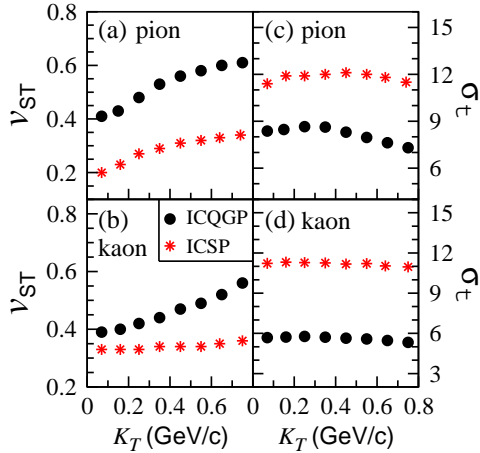}
\caption{\label{fig:svtt} (Color online) Left panels: the transverse
velocities of the particle-emitting sources as functions of $K_T$.
Right panels: the standard deviations of time of the particles-emitting
sources. }
\end{figure}

Figure \ref{fig:svtt} (a) and (b) show the transverse velocities of
the pion- and kaon-emitting sources as functions of the pair transverse
momenta $K_T$.  Figure \ref{fig:svtt} (c) and (d) show the standard
deviations of time, $\sigma_t=\sqrt{\langle\,(t - \langle t \rangle)^2\,
\rangle}$, of the particle-emitting sources.  One can see that the
transverse velocities of the pion and kaon sources are smaller for the
system initially at the soft point (ICSP) than those for the system
initially in the QGP (ICQGP).  The standard deviations of time enhance
very much for the sources for ICSP.

\begin{figure}
\vspace*{-5mm}
\hspace*{4mm}
\includegraphics[angle=0,scale=0.60]{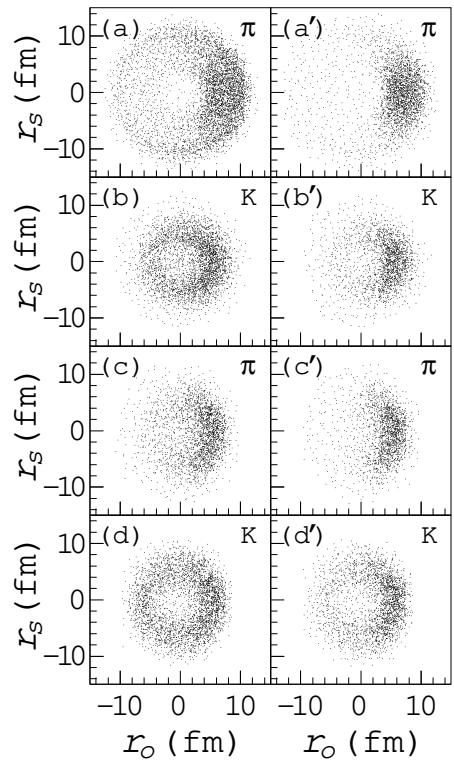}
\vspace*{-15mm}
\caption{\label{fig:densxy} The distributions of the source coordinates
projected on the transverse out-side plane.  The left and right panels are
for the smaller and larger particle pair momenta $K_T<300$ MeV/c and $K_T>
300$ MeV/c.  The upper four panels are for ICQGP.  The lower four panels
are for ICSP. }
\end{figure}

In HBT interferometry, the source HBT radii are related to the enhancements
of the correlation functions at small relative momenta.  For an evolving
source, the source expansion leads to a correlation between the particle-emitting
coordinate and momentum.  It may decrease the transverse emission region for the
particle pairs with small relative momenta and large $K_T$.  This effect is more
important in the direction of the transverse momentum of the pair (out direction),
which is boosted by the source expansion.  Additionally, the source opacity, due
to the absorptions for the particles propagating through the center of source
(in which the temperatures are higher than the hadronization temperature) and by
the multiple scattering among the particles in the source, may lead to a shell
emission.  This will increase the effect of the decrease of emission region
for expanding sources.  In Fig \ref{fig:densxy}, we show the distributions
of the source coordinates projected on the transverse out-side plane, for
the particles with the smaller pair momenta $K_T<300$ MeV/c (left panels)
and the larger pair momentum $K_T>300$ MeV/c (right panels).  The upper
four panels are for the system initially in the QGP (ICQGP).  The lower
four panels are for the system initially at the soft point (ICSP).
For $K_T>300$ MeV/c, the distributions of the source coordinates are more
concentrated in $r_o>0$ regions.  For $K_T<300$ MeV/c, the annular
distributions for kaon indicate that the sources are almost transparent
for the kaons emitted later.  We will see it is that the source expansion,
lifetime ($\,\sim \sigma_t$), and particle absorptions lead to the differences
of the transverse HBT radii $R_{\rm out}$ for the two kinds of sources for
ICQGP and ICSP.

\subsection{The effects of source expansion and lifetime on transverse
HBT radii}

In HBT interferometry, the difference of the transverse HBT radii in out
and side directions includes the important information on the source
expansion and lifetime.  $R_{\rm out}^2 -R_{\rm side}^2$ is given by
\cite{Her95,Cha95,Wie99}
\begin{equation}
\label{Ros}
R_{\rm out}^2 - R_{\rm side}^2 = \langle\,({\tilde r_o} - v_{_{\rm KT}}
{\tilde t\,})^2 \rangle - \langle\, {\tilde r_s}^2 \rangle\,,
\end{equation}
where $\langle\, \cdots \rangle$ denotes the average for the space-time
coordinates of the source, $\tilde r_o$ and $\tilde r_s$ are the biases
of the source spatial coordinates related to their average values in the
out and side directions, $\tilde t$ is the bias of the source time
coordinate related to its average, and $v_{_{\rm KT}}$ is the transverse
velocity of the particle pair.

In order to examine the effects of source expansion and lifetime on
the transverse HBT radii $R_{\rm out}$ and $R_{\rm side}$, we investigate
next the two-pion interferometry for the simple sources with a constant
temperature 100 MeV and the Gaussian space-time distributions as
\begin{eqnarray}
\frac{dN}{d^3r dt} &\propto& \exp \bigg( -\frac{x^2+y^2}{2R_T^2}
-\frac{z^2}{2R_L^2} -\frac{t^2}{2\tau^2}\bigg)\,,\nonumber\\
& &R_1 \le \sqrt{x^2+y^2+z^2} \le R_2\,.
\end{eqnarray}
We take $R_T=R_L=5$ fm, $R_2=10$ fm, and assume that the sources have
the radial velocity
\begin{equation}
v_r=v_0 \frac{r}{R_2}.
\end{equation}
Here $\tau$, $R_1$, and $v_0$ are three free parameters.  We taken $\tau
=6$ and 12 fm/c for the sources with shorter and longer lifetimes.  For
a shell source $R_1$ is taken to be 5 fm.  For static and expanding
sources, $v_0$ is taken to be 0 and 0.8 respectively.  Because there are
not correlations between spatial coordinates and time for these sources,
Eq. (\ref{Ros}) reduces to
\begin{equation}
\label{Ros1}
R_{\rm out}^2 -R_{\rm side}^2 =\langle\,{\tilde r_o}^2\rangle -\langle\,
{\tilde r_s}^2 \rangle + \sigma_t^2 v_{_{\rm KT}}^2\,,
\end{equation}
where $\sigma^2_t=\langle\,(t - \langle t \rangle)^2\rangle =(\pi-2)\tau^2
/\pi \approx 0.363 \tau^2$.

Because of source expansion and opacity the difference of the variances
in out and side directions, $\langle\,{\tilde r_o}^2 \rangle -\langle\,
{\tilde r_s}^2 \rangle$, is not zero even for the source with transverse
symmetry.  It is negative and decrease with $K_T$.  On the other hand,
the right third term in Eq. (\ref{Ros1}), $\sigma_t^2 v_{_{\rm KT}}^2$,
is positive.  It increases with $K_T$ and becomes important when the source
lifetime $\tau$ increases.

\begin{figure}
\includegraphics[angle=0,scale=0.8]{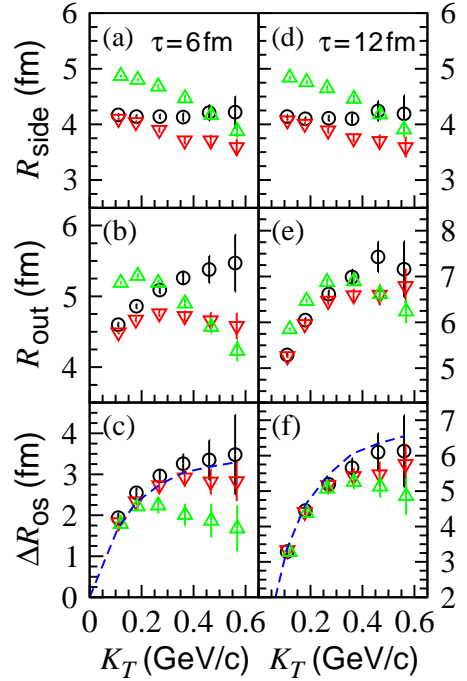}
\caption{\label{fig:HBTrgs} (Color online) The transverse HBT radii and
$\Delta R_{\rm os}=\sqrt{R^2_{\rm out} -R^2_{\rm side}}$ for the static
Gaussian source (circle), expanding Gaussian source (triangle down), and
expanding shell source (triangle up).  The dashed lines are the results
of $\sigma_t v_{_{\rm KT}}$. }
\end{figure}

\begin{figure}
\vspace*{-3mm}
\includegraphics[angle=0,scale=0.50]{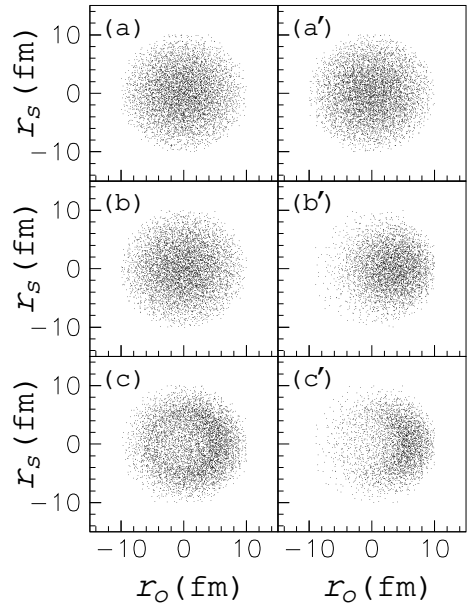}
\vspace*{-20mm}
\caption{\label{fig:rors_gs} The distributions of the source coordinates
projected on $r_o - r_s$ plane.  The panels (a), (b), and (c) are for the
static Gaussian, expanding Gaussian, and expanding shell sources for
$K_T<300$ MeV/c.  The panels (a$'$), (b$'$), and (c$'$) are for the
static Gaussian, expanding Gaussian, and expanding shell sources for
$K_T>300$ MeV/c. }
\end{figure}

In Fig. \ref{fig:HBTrgs}, we show the transverse HBT radii $R_{\rm out}$
and $R_{\rm side}$ and $\Delta R_{\rm os}=\sqrt{R^2_{\rm out} -R^2_{\rm
side}}$ for the sources with $\tau=$ 6 and 12 fm.  The symbols $\circ$,
{\raisebox{-1.3mm}{\textsuperscript{$\nabla$}}}, and {\raisebox{-1.3mm}
{\textsuperscript{$\triangle$}}} are for the static Gaussian source ($v_0=0$,
$R_1=0$ fm), expanding Gaussian source ($v_0=0.8$, $R_1=0$ fm), and expanding
shell source ($v_0=0.8$, $R_1=5$ fm).  The dashed lines in the bottom panels
are the results of $\sigma_t v_{_{\rm KT}}$ ($v_{_{\rm KT}}=K_T/E_K$, $E_K=
(E_1+E_2)/2$.)  In Fig. \ref{fig:rors_gs}, we show the distributions of the
source coordinates projected on $r_o-r_s$ plane.  The panels (a), (b), and
(c) are for the static Gaussian source, expanding Gaussian source, and
expanding shell source for the smaller pion pair momentum $K_T<300$ MeV/c.
The panels (a$'$), (b$'$), and (c$'$) are for the static Gaussian source,
expanding Gaussian source, and expanding shell source for $K_T>300$ MeV/c.
For the static sources, the results of $R_{\rm side}$ are almost a constant
and $R_{\rm out}$ increases with $K_T$.  Because there is not the effect of
source expansion, $\langle\,{\tilde r_o}^2\rangle = \langle\,{\tilde r_s}^2
\rangle$, and the results of $\Delta R_{\rm os}$ are consistent with those
of $\sigma_t v_{_{\rm KT}}$.  For the expanding and shell-emitting sources,
the source expansion and shell emission change the distributions of the
source coordinates.  It leads to the decreases of $R_{\rm side}$ with $K_T$.
Although $R_{\rm out}$ increases with $K_T$ in the small $K_T$ region for
the sources with larger lifetime $\tau$, this increase will be counteracted
at large $K_T$ by the effects of the source expansion and shell emission.
In these cases, the results of $\Delta R_{\rm os}$ are smaller than the
values of $\sigma_t v_{_{\rm KT}}$ at larger $K_T$.
From Fig. \ref{fig:rors_gs} one can see directly that the coordinate distributions
of the static sources for the smaller and larger pion pair momenta are almost
the same.  However, the coordinate distributions of the expanding sources for
$K_T>300$ MeV/c are more concentrated in $r_o>0$ regions as compared to the
corresponding distributions for $K_T<300$ MeV/c.

For hydrodynamic sources, there are also correlations between source spatial
coordinates and time.  We will discuss the effect of the correlation between
$r_o$ and $t$ on $\Delta R_{\rm os}$ in next subsection.

\subsection{Characteristic quantity for long source lifetime for ICSP}

Because of the correlation between source spatial coordinate $r_o$ and time
for hydrodynamic sources, Eq. (\ref{Ros}) becomes
\begin{equation}
\label{Ros2}
R_{\rm out}^2 -R_{\rm side}^2 =\langle\,{\tilde r_o}^2\rangle -\langle\,
{\tilde r_s}^2 \rangle + \sigma_t^2 v_{_{\rm KT}}^2 -2\langle\,{\tilde r_o}\,
{\tilde t}\, \rangle\, v_{_{\rm KT}}\,.
\end{equation}
For positive or negative $\langle\,{\tilde r_o}\,{\tilde t}\, \rangle$, the
right last term in Eq. (\ref{Ros2}) will decrease or increase $\Delta R_{\rm os}$
with $K_T$ increase.

\begin{figure}
\vspace*{-3mm}
\includegraphics[angle=0,scale=0.55]{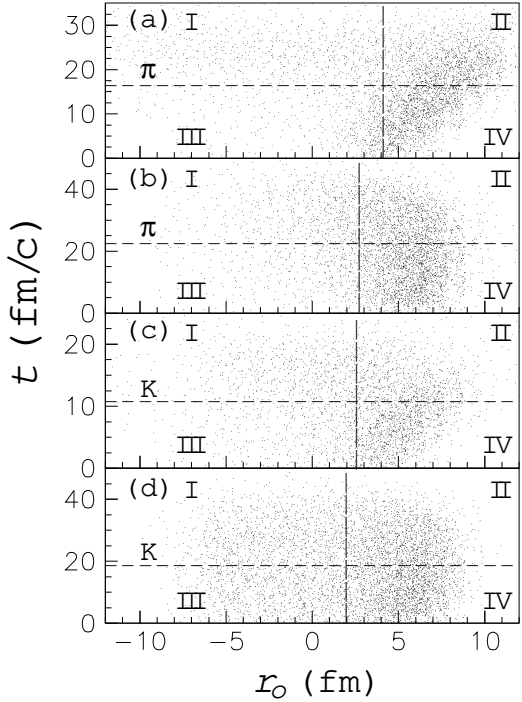}
\vspace*{-15mm}
\caption{\label{fig:densxt} The distributions of the space-time coordinates
of source points projected on $r_o - t$ plane for ICQGP ((a) and (c))
and ICSP ((b) and (d)).  The dashed lines are for the average values of
$r_o$ and $t$. }
\end{figure}

In Fig. \ref{fig:densxt} we show the distributions of the space-time
coordinates of source points projected on $r_o - t$ plane for ICQGP
(panels (a) and (c)) and ICSP (panels (b) and (d)).  The dashed lines are
for the average values of $r_o$ and $t$, which divide the plane into four
regions.  In regions I and IV, $\langle\,{\tilde r_o}\,{\tilde t}\,\rangle
<0$.  In regions II and III, $\langle\,{\tilde r_o}\,{\tilde t}\,\rangle
>0$.  For pion, because of particle absorption the distributions in region
III are less than those in region I.  So the values of ${\tilde r_o}
{\tilde t}$ averaging over all $r_o<\langle\,r_o\,\rangle$ regions (I and
III) are negative.  In $r_o>\langle\,r_o\,\rangle$ regions, the distribution
for the pion source for ICQGP (panel (a)) is much different from that for
ICSP (panel (b)) because of the larger transverse expansion of source for
ICQGP.  In region II of the panel (a), the wider distribution for the pion
source for ICQGP leads to a greater contribution to $\langle \, {\tilde r_o}\,
{\tilde t}\,\rangle$.  So the value of ${\tilde r_o} {\tilde t}$ averaging over
all $r_o >\langle\,r_o\,\rangle$ regions (II and IV) is positive for ICQGP.
For ICSP, the distribution for pion source in region II is less than that
in region IV.  The value of ${\tilde r_o} {\tilde t}$ averaging over all
$r_o >\langle\,r_o\,\rangle$ regions is also negative for ICSP.  The
calculations indicate that for pion $\langle\,{\tilde r_o}\,{\tilde t}\,
\rangle=$0.16 and $-$8.30 fm$^2$/c for ICQGP and ICSP, which are consistent
with the above discussions.  For kaon, because of the high transparency
and low transverse expansion of sources, the distributions are
approximately symmetric about $\langle\, t\, \rangle$.  The values of
$\langle\, {\tilde r_o}\,{\tilde t}\,\rangle$ are small for the kaon
sources.

\begin{figure}
\includegraphics[angle=0,scale=0.8]{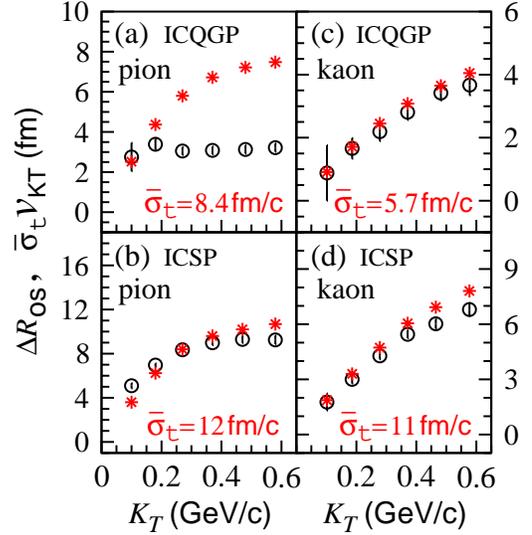}
\caption{\label{fig:dros} (Color online) The results of $\Delta R_{\rm os}$
($\circ$) and $\bar \sigma_t v_{_{\rm KT}}$ ($\ast$) for the hydrodynamic
sources for ICQGP and ICSP. }
\end{figure}

In Fig. \ref{fig:dros}, we show the results of $\Delta R_{\rm os}$ (symbols
$\circ$) and the products $\bar \sigma_t v_{_{\rm KT}}$ (symbols $\ast$) of
the average emission durations $\bar\sigma_t$ of the particles and the pair
transverse velocities $v_{_{\rm KT}}$ for the hydrodynamic sources for ICQGP
and ICSP.  It can be seen that except for the pion results in panel (a), the
results of the $\Delta R_{\rm os}$ and $\bar \sigma_t v_{_{\rm KT}}$ are
almost consistent.

Inspired by the consistences of the results of $\Delta R_{\rm os}$ and
$\bar \sigma_t v_{_{\rm KT}}$ for ICSP, we introduce the quantity
\begin{equation}
\tilde \sigma = \frac{\Delta R_{\rm os}}{v_{_{\rm KT}}}=\frac{\sqrt{
R_{\rm out}^2 -R_{\rm side}^2}}{{K_T}/{E_K}},
\end{equation}
to describe the character of the long lifetime of the sources for ICSP.
It is an experimental observable.

\begin{figure}
\vspace*{3mm}
\includegraphics[angle=0,scale=0.8]{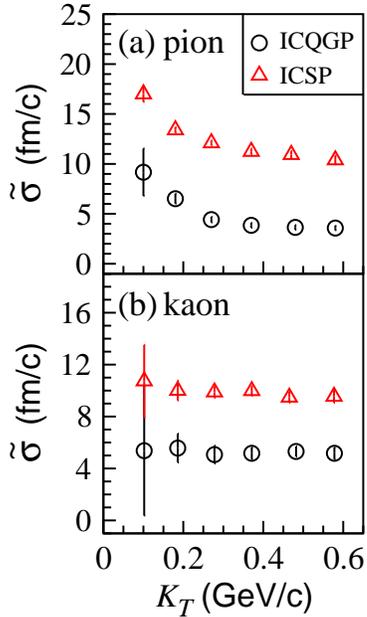}
\caption{\label{fig:sigma} (Color online) The results of $\tilde \sigma$
for the hydrodynamic sources for ICQGP and ICSP.}
\end{figure}

In Fig. \ref{fig:sigma} we show the results of $\tilde \sigma$ for pion
and kaon for the hydrodynamic sources for ICQGP and ICSP.  The larger
values of $\tilde \sigma$ for the soft point of the first-order phase
transition are observed in both of the pion and kaon interferometry
measurements.  The $\tilde \sigma$ values for the pion source for ICQGP
are much smaller than the average value $\bar\sigma_t=$8.4 fm/c at large
$K_T$, because of the large transverse velocities of the source and the
positive values of $\langle\,{\tilde r_o}\,{\tilde t}\,\rangle$.  In this
case $\tilde\sigma$ cannot reflect the real lifetime of the source.  At
small $K_T$, the larger values of $\tilde\sigma$ for the pion source for
ICSP are due to the large negative values of $\langle\,{\tilde r_o}\,
{\tilde t}\,\rangle$ as well as the small transverse velocities of the
source in this case.
The errors of $\tilde \sigma$ exhibited in Fig. \ref{fig:sigma} are only
from the statistic errors of $R_{\rm out}$ and $R_{\rm side}$ related to
the HBT parametrized fits.  In high energy heavy ion collisions, there
are other effects which may bring uncertainty to the observable,
for example the non-equilibrium dynamics during the decay of resonances
after the hadronization.  Further investigations on these effects will
be of interest.

\section{Summary and conclusions}

We investigate the two-particle HBT interferometry for the hydrodynamic
particle-emitting sources which undergo the first-order phase transition
from the quark-gluon plasma with finite baryon chemical potentials to
hadron resonance gas.  The effects of source expansion, lifetime, and
particle absorption on the HBT radii are examined.  For pion, the large
transverse expansion of the source for ICQGP decreases the HBT radii
$R_{\rm out}$ and $R_{\rm side}$ at large transverse momentum of
particle pair $K_T$.  The source has a long lifetime and small expansion
when the system is initially located at the boundary between the mixed
phase and the QGP (soft point).  In this case, the difference between
the transverse HBT radii $R_{\rm out}$ and $R_{\rm side}$ increases with
$K_T$ significantly.  The ratio of $\sqrt{R_{\rm out}^2-R_{\rm side}^2}$
to the transverse velocity of the particle pair $v_{_{\rm KT}}$, $\tilde
\sigma$, is an observable for probing the long lifetime of the source for
the soft point of the first-order phase transition.  As compared to pion
HBT interferometry kaon HBT interferometry may present more clearly the
source space-time geometry at the emission.  The larger values of $\tilde
\sigma$ for the soft point of the first-order phase transition can be
observed both by two-pion and two-kaon HBT measurements.
Further investigations on other effects on the observable will be of 
interest. \\[2ex]

\begin{acknowledgments}
The authors would like to thank Dr. C. Y. Wong for helpful
discussions.  This research was supported by the National Natural
Science Foundation of China under grant 11075027.
\end{acknowledgments}

\end{document}